\documentclass[a4paper,12pt]{article}
\usepackage{epsfig,amssymb,amsmath}
\usepackage{graphicx}
\usepackage{amsfonts}

\begin{document}
%

\newcommand{\be}{\begin{equation}}
\newcommand{\ee}{\end{equation}}
\newcommand{\bea}{\begin{eqnarray}}
\newcommand{\eea}{\end{eqnarray}}
\newcommand{\bean}{\begin{eqnarray*}}
\newcommand{\eean}{\end{eqnarray*}}
\font\upright=cmu10 scaled\magstep1
\font\sans=cmss12
\newcommand{\ssf}{\sans}
\newcommand{\stroke}{\vrule height8pt width0.4pt depth-0.1pt}
\newcommand{\Z}{\hbox{\upright\rlap{\ssf Z}\kern 2.7pt {\ssf Z}}}
\newcommand{\ZZ}{\Z\hskip -10pt \Z_2}
\newcommand{\C}{{\rlap{\upright\rlap{C}\kern 3.8pt\stroke}\phantom{C}}}
\newcommand{\R}{\hbox{\upright\rlap{I}\kern 1.7pt R}}
\newcommand{\HH}{\hbox{\upright\rlap{I}\kern 1.7pt H}}
\newcommand{\CP}{\hbox{\C{\upright\rlap{I}\kern 1.5pt P}}}
\newcommand{\identity}{{\upright\rlap{1}\kern 2.0pt 1}}
\newcommand{\half}{\frac{1}{2}}
\newcommand{\quart}{\frac{1}{4}}
\newcommand{\pr}{\partial}
\newcommand{\bm}{\boldmath}
\newcommand{\I}{{\cal I}} 
\newcommand{\M}{{\cal M}}
\newcommand{\N}{{\cal N}}
\newcommand{\e}{\varepsilon}
\newcommand{\ep}{\epsilon}
\newcommand{\bep}{\mbox{\boldmath $\varepsilon$}}
\newcommand{\Oh}{{\rm O}}
\newcommand{\n}{{\bf n}}
\newcommand{\x}{{\bf x}}
\newcommand{\y}{{\bf y}}
\newcommand{\X}{{\bf X}}
\newcommand{\Y}{{\bf Y}}
\newcommand{\z}{{\bar z}}
\newcommand{\w}{{\bar w}}
\newcommand{\tT}{{\tilde T}}
\newcommand{\tX}{{\tilde\X}}

\thispagestyle{empty}
\rightline{DAMTP-2014-35}
\vskip 5em
\begin{center}
{\bf \Large Rational Relativistic Collisions} 
\\[15mm]

{\bf \Large N.~S. Manton\footnote{email: N.S.Manton@damtp.cam.ac.uk}} \\[20pt]

\vskip 3em
{\it 
Department of Applied Mathematics and Theoretical Physics,\\
University of Cambridge, \\
Wilberforce Road, Cambridge CB3 0WA, U.K.}
\vspace{20mm}

\abstract{}
If two point particles collide relativistically in one dimension, and
the masses, velocities and gamma factors of the incoming particles 
are rational numbers, then the velocities and gamma factors of the 
outgoing particles are rational. Numerous examples can be found using 
Pythagorean triples. At all velocities, the collision results in a 
Lorentzian reflection of the 2-momenta of the particles.

\end{center}

\vskip 100pt
\leftline{Keywords:} 
\leftline{1-d Relativistic collisions, Rational velocities,
Pythagorean triples} 

\vskip 5pt

\vfill
\newpage
\setcounter{page}{1}
\renewcommand{\thefootnote}{\arabic{footnote}}


\section{Introduction} 
\vspace{3mm}

One of the most interesting problems in elementary relativity is to find the 
outcome of a 1-dimensional elastic collision of two particles \cite{LL,Woo}. 
The masses and incoming velocities of the particles are assumed known. 
The masses are unchanged in the collision, and the problem is to find the 
outgoing velocities. The physical principle governing the collision is 
relativistic 2-momentum conservation, combining the requirements of 
relativistic energy and momentum conservation. This 
gives two equations for the outgoing velocities. The solution
is different from what one gets using non-relativistic, Newtonian
energy and momentum conservation. Experiments show that for particles colliding
at speeds comparable to the speed of light, the relativistic 
results are correct. We set the speed of light to be $1$.

The conservation equations are rather intimidating, because of 
gamma factors in each term. For velocity $u$, the gamma factor 
is $\gamma(u) = (1 - u^2)^{-\half}$. It is easy to get bogged down 
trying to eliminate the square roots. One might expect that having done this,
the resulting algebraic equations would be quadratic or
quartic, and their solution would involve at least one square root. 
However, the text-book solution shows that the 
energy of one of the outgoing particles is a simple rational 
function of the incoming energies and momenta.
Here we go further, and show that if the mass and velocity data of the incoming
particles are all rational, in a sense that we will clarify, then
the outgoing data are rational too. 

It is easy to see that in a non-relativistic elastic collision, 
rational incoming data implies rational outgoing data. Suppose the 
incoming particles have masses $m_1$ and $m_2$, and velocities 
$u_1$ and $u_2$, and suppose the outgoing velocities 
are $v_1$ and $v_2$.  Newtonian energy and momentum conservation require that
\be
m_1 u_1^2 + m_2 u_2^2 = m_1 v_1^2 + m_2 v_2^2  
\label{Newtenercons}
\ee
and
\be
m_1 u_1 + m_2 u_2 = m_1 v_1 + m_2 v_2 \,,  
\label{Newtmomcons}
\ee
where we have eliminated the $\half$ in the kinetic
energies. Superficially, it appears that in solving for $v_1$ and
$v_2$, we will encounter square roots.

It is convenient to work in an inertial frame where the second 
particle is initially at rest, so $u_2 = 0$. Then,
to find $v_2$ we take the terms involving $v_2$ to the left hand
side, and eliminate $v_1$ by squaring the second equation and
subtracting $m_1$ times the first. The result is a quadratic
equation 
\be
(m_1 + m_2)v_2^2 - 2m_1u_1v_2 = 0 \,,
\label{v2quad}
\ee
but there is no term independent of $v_2$. One solution is therefore 
$v_2 = 0$, and this is because one possible outcome, consistent with 
energy and momentum conservation, is that the particles miss and retain their
initial velocities. The quadratic equation (\ref{v2quad}) therefore 
reduces to a linear equation, with solution
\be
v_2 = \frac{2m_1}{m_1 + m_2}u_1 \,.
\label{v2soln}
\ee 

The solution for general $u_2$ is found by performing a Galilean 
boost to make $u_2=0$, then determining $v_2$ using eq.(\ref{v2soln}),
and finally reversing the boost. $v_1$ is then easily found using 
eq.(\ref{Newtmomcons}). The final, rational expressions for $v_1$ and $v_2$ are
very similar, because of particle exchange symmetry. They are
\be
v_1 = \frac{m_1 - m_2}{m_1 + m_2}u_1 + \frac{2m_2}{m_1 + m_2}u_2 \,,
\quad v_2 = \frac{2m_1}{m_1 + m_2}u_1 + \frac{m_2 - m_1}{m_1 + m_2}u_2 \,. 
\ee
The outgoing velocities are therefore rational if the masses and
incoming velocities are all rational.

Before looking at the details of relativistic collisions,
we recall certain aspects of the rational geometry of a unit circle
and hyperbola, and the role of Pythagorean triples.  

\vspace{4mm}

\section{Rational points on a circle and hyperbola}
\vspace{3mm}

It will be helpful to recall three parametrisations of a circle, and
the related parametrisations of a hyperbola. Consider the unit circle
\be
x^2 + y^2 = 1 \,.
\ee
The first parametrisation, using just $y$, is $(x,y) = ((1 - y^2)^\half, y)$. 
$y$ is between $-1$ and $1$, and both square roots are needed. The second 
is the trigonometrical parametrisation $(x,y) = (\cos\theta, \, \sin\theta)$, 
with $\theta$ between $-\pi$ and $\pi$. Finally, there is the rational 
parametrisation
\be
x = \frac{1 - t^2}{1 + t^2} \,, \quad y = \frac{2t}{1 + t^2} \,,
\ee
with $t$ between $-\infty$ and $\infty$. The relation of $t$ to
$\theta$ is $t = \tan\half\theta$.

Rational points on the circle are those with both $x$ and $y$ rational, and
therefore $\cos\theta$ and $\sin\theta$ both rational. They are the
points for which $t$ is rational, so they are dense on the circle. 
If $t = \frac{N}{M}$, with $M$ and $N$ integers, then
\be
x = \cos\theta = \frac{M^2 - N^2}{M^2 + N^2} \,, \quad 
y = \sin\theta = \frac{2MN}{M^2 + N^2} \,.
\ee
A rational point is therefore related to a Pythagorean triple, a triple
of integers $(A,B,C) = (M^2 - N^2, \, 2MN, \, M^2 + N^2)$, for which $A^2 +
B^2 = C^2$.
Pythagorean triples that differ by a constant integer multiple, e.g.
$(3,4,5)$ and $(6,8,10)$, correspond to the same point on the circle,
so rational points are uniquely related to primitive Pythagorean triples.
Examples of rational points are
\be
(x,y) = \left(\frac{3}{5}, \frac{4}{5} \right) \quad {\rm and} \quad
(x,y) = \left(\frac{5}{13}, \frac{12}{13} \right) \,,
\ee
obtained from the triples $(3,4,5)$ with $(M,N) = (2,1)$, and 
$(5,12,13)$ with $(M,N) = (3,2)$.

The rational points form a group, isomorphic to the rational subgroup 
of SO(2). This is the group of matrices
\be
\left(\begin{array}{cc}
\cos\theta &  -\sin\theta \\
\sin\theta & \cos\theta 
\end{array}\right) \,,
\ee
with $\cos\theta$ and $\sin\theta$ both rational. Each rational point
$(x,y)$ (written as a column vector) is obtained by acting with a rational 
rotation matrix on the point $(1,0)$. The group law is addition of 
angles, even though the angles are not rational.

A similar analysis can be applied to the hyperbola 
\be
x^2 - y^2 = 1 \,.
\ee
We only consider the branch with $x \ge 1$ (having in mind the
positive energy branch of the relativistic mass-shell condition).
Two parametrisations are $(x,y) = ((1 + y^2)^\half, y)$, with $y$
between $-\infty$ and $\infty$ and the square root positive, and the 
parametrisation $(x,y) = (\cosh\theta, \, \sinh\theta)$, using the 
hyperbolic functions, with $\theta$ between $-\infty$ and $\infty$. 
Finally, there is the rational parametrisation
\be
x = \frac{1 + t^2}{1 - t^2} \,, \quad y = \frac{2t}{1 - t^2} \,,
\ee
with $t$ between $-1$ and $1$. The relation of $t$ to $\theta$ is 
$t = \tanh\half\theta$. There is a second rational parametrisation
\be
x = \frac{1 + s^2}{2s} \,, \quad y = \frac{1 - s^2}{2s} \,,
\ee
but this is rationally obtained from the first by setting $t = \frac{1-s}{1+s}$.

By setting $t = \frac{N}{M}$, with $M^2 > N^2$, one obtains the rational point 
\be
x = \cosh\theta = \frac{M^2 + N^2}{M^2 - N^2} \,, \quad 
y = \sinh\theta = \frac{2MN}{M^2 - N^2} \,.
\label{hypPythpts}
\ee
This is again related to the Pythagorean triple $(M^2 - N^2, \, 2MN,
\, M^2 + N^2)$. Examples of rational points are 
$(x,y) = (\frac{5}{3}, \frac{4}{3})$ and $(\frac{5}{4}, \frac{3}{4})$, 
related to the triples with $(M,N) = (2,1)$ and $(3,1)$. 

The rational points are dense on the hyperbola. They form a
group isomorphic to the rational subgroup of SO(1,1), the
group of matrices
\be
\left(\begin{array}{cc}
\cosh\theta &  \sinh\theta \\
\sinh\theta & \cosh\theta 
\end{array}\right)
\ee
with $\cosh\theta$ and $\sinh\theta$ both rational. Each rational point
is obtained by acting with a matrix of this type on the point
$(1,0)$.

The rational points can also be understood by writing the
equation for the hyperbola as $(x-y)(x+y) = 1$. The rational solutions
are
\be
(x-y, \, x+y) = \left(\frac{M-N}{M+N}, \, \frac{M+N}{M-N}\right) \,,
\ee
and adding and subtracting these gives the formulae (\ref{hypPythpts}).

\vspace{4mm}

\section{Rational masses and 2-velocities}
\vspace{3mm}

We next recall the relativistic energy and momentum of a particle
with mass $m$ and velocity $u$ moving in one dimension. Associated 
with the velocity is a (positive) gamma factor $\gamma(u) = (1 - u^2)^{-\half}$.

The 2-velocity $U$ of the particle is
\be
U = (\gamma(u), \gamma(u)u) \,.
\ee
This is a 2-vector for which the Lorentzian inner product $U \cdot
U = (\gamma(u))^2 - (\gamma(u)u)^2$ is unity, which follows 
immediately from the definition of $\gamma(u)$. So $U$ is on 
the hyperbola $x^2 - y^2 = 1$ that we discussed in the last section, 
and on the branch with $x$ positive. We can use the parametrisation 
in terms of hyperbolic functions and set $\gamma(u) = \cosh\theta$ 
and $\gamma(u)u = \sinh\theta$; then the 2-velocity is 
$U = (\cosh\theta, \, \sinh\theta)$, and $u = \tanh\theta$. $\theta$ 
is called the rapidity of the particle.

We say a 2-velocity is rational if $\gamma(u)$ and $\gamma(u)u$ are
rational, so $u$ is too. Equivalently, 
$\sinh\theta$, $\cosh\theta$ and $\tanh\theta$ are all rational, and
this requires $t = \tanh \half\theta$ to be rational. The 
velocities $u$ giving rational 2-velocities are dense between $-1$ and $1$.
We deduce from eqs.(\ref{hypPythpts}) that for a rational 2-velocity, 
$u$ and $\gamma(u)$ can be expressed in terms of the integers $(M,N)$ 
of a Pythagorean triple, with $M^2 > N^2$, as
\be
u = \frac{2MN}{M^2 + N^2} \,, \quad 
\gamma(u) = \frac{M^2 + N^2}{M^2 - N^2} \,,
\ee
so $\gamma(u)u = \frac{2MN}{M^2 - N^2}$. The examples $(M,N) = (2,1)$,
$(3,1)$ and $(3,2)$ give, respectively, $(u, \gamma(u)) =
\left(\frac{4}{5}, \frac{5}{3}\right)$, $\left(\frac{3}{5}, 
\frac{5}{4}\right)$ and $\left(\frac{12}{13},\frac{13}{5}\right)$. 

The relativistic 2-momentum of a particle with mass $m$ and 2-velocity 
$U$ is $mU$. Its components are written as 
\be
mU = (m \gamma(u), \, m \gamma(u)u) \,,
\ee
where the first component is the relativistic energy and the second 
is the relativistic, spatial momentum. These satisfy the mass-shell condition 
$(m \gamma(u))^2 - (m \gamma(u)u)^2 = m^2$. In terms of the particle 
mass $m$ and rapidity $\theta$,
\be
mU = (m \cosh\theta, \, m \sinh\theta) \,.
\ee
The 2-momentum is rational if both the mass and 2-velocity are rational.

\vspace{4mm} 

\section{Relativistic collisions}
\vspace{3mm}

We can now investigate the main problem, the relativistic elastic 
collision of two particles with masses $m_1$ and $m_2$, and given 
initial velocities $u_1$ and $u_2$. We use the hyperbolic parametrisation, 
$u_1 = \tanh\phi_1$ and $u_2 = \tanh\phi_2$. The initial 2-momenta of
the particles are 
$m_1U_1 = (m_1\cosh\phi_1, \, m_1\sinh\phi_1)$ and 
$m_2U_2 = (m_2\cosh\phi_2, \, m_2\sinh\phi_2)$. 
The final, unknown velocities are denoted $v_1 = \tanh\chi_1$ and 
$v_2 = \tanh\chi_2$, and the corresponding final 2-momenta are 
$m_1V_1 = (m_1\cosh\chi_1, \, m_1\sinh\chi_1)$ and 
$m_2V_2 = (m_2\cosh\chi_2, \, m_2\sinh\chi_2)$. The initial data are
rational if $m_1, m_2, \cosh\phi_1, \sinh\phi_1, \cosh\phi_2$
and $\sinh\phi_2$ are all rational.

The physical principle determining $v_1$ and $v_2$ is the conservation
of total 2-momentum, 
\be
m_1U_1 + m_2U_2 = m_1V_1 + m_2V_2 \,.
\label{relMomcons}
\ee
In components,
\bea
m_1\cosh\phi_1 + m_2\cosh\phi_2 &=& m_1\cosh\chi_1 + m_2\cosh\chi_2
\,, \nonumber \\
m_1\sinh\phi_1 + m_2\sinh\phi_2 &=& m_1\sinh\chi_1 + m_2\sinh\chi_2 \,,
\label{rel2momcons}
\eea
the equations of relativistic energy and momentum conservation. These
are the equations with intimidating square roots if one writes them
using explicit gamma factors, that is, with $\cosh \phi_1 = (1 -
u_1^2)^{-\half}$, $\sinh \phi_1 = (1 - u_1^2)^{-\half} u_1$, etc.

It is convenient, as in the Newtonian case, to first assume that the second
particle is initially at rest. So $\phi_2 = 0$, and $\cosh\phi_2 = 1$,
$\sinh\phi_2 = 0$. To find $\cosh\chi_2$ and $\sinh\chi_2$, we 
eliminate $\chi_1$ by taking the terms involving 
$\chi_2$ to the left hand side, then square the component equations, and
take their difference. Using the identity $\cosh^2\theta 
- \sinh^2\theta = 1$, there remains the single equation
\be
(m_1\cosh\phi_1 + m_2 - m_2\cosh\chi_2)^2 
- (m_1\sinh\phi_1 - m_2\sinh\chi_2)^2 = m_1^2 \,.
\label{elimchi1}
\ee
After expanding out, and using the identity again, eq.(\ref{elimchi1}) 
simplifies to 
\be
(m_1\cosh\phi_1 + m_2)(\cosh\chi_2 - 1) = m_1\sinh\phi_1 \sinh\chi_2 \,,
\label{lincoshsinh}
\ee
an equation linear in $\cosh\chi_2$ and $\sinh\chi_2$. Squaring 
this, and using the identity once more, gives 
\bea
&&\bigl((m_1\cosh\phi_1 + m_2)^2 - m_1^2\sinh^2\phi_1\bigr)\cosh^2\chi_2
\nonumber \\
&& \qquad - 2(m_1\cosh\phi_1 + m_2)^2\cosh\chi_2 \nonumber \\
&& \qquad\qquad + \bigl((m_1\cosh\phi_1 + m_2)^2 
+ m_1^2\sinh^2\phi_1\bigr) = 0 \,, 
\eea
a quadratic equation for $\cosh\chi_2$ alone.

As in the Newtonian case, one solution is known. It
is $\cosh\chi_2 = 1$, corresponding to the particles
missing each other. The other solution is therefore rational in the
coefficients and is
\be
\cosh\chi_2 = \frac{(m_1\cosh\phi_1 + m_2)^2 + m_1^2\sinh^2\phi_1}
{(m_1\cosh\phi_1 + m_2)^2 - m_1^2\sinh^2\phi_1} \,.
\ee
It follows from eq.(\ref{lincoshsinh}) that
\be
\sinh\chi_2 = \frac{2(m_1\cosh\phi_1 + m_2)m_1\sinh\phi_1}
{(m_1\cosh\phi_1 + m_2)^2 - m_1^2\sinh^2\phi_1} \,.
\ee
By cancelling a factor of $m_1^2$, these formulae simplify to
\be
\cosh\chi_2 = \frac{(\cosh\phi_1 + \frac{m_2}{m_1})^2 + \sinh^2\phi_1}
{(\cosh\phi_1 + \frac{m_2}{m_1})^2 - \sinh^2\phi_1} \,, \
\sinh\chi_2 = \frac{2(\cosh\phi_1 + \frac{m_2}{m_1})\sinh\phi_1}
{(\cosh\phi_1 + \frac{m_2}{m_1})^2 - \sinh^2\phi_1} \,,
\label{chi2}
\ee
showing that the outcome of a collision only depends on the mass ratio 
of the colliding particles. The outgoing velocity of particle 2 is
\be
v_2 = \tanh\chi_2 = \frac{2(\cosh\phi_1 + \frac{m_2}{m_1})\sinh\phi_1}
{(\cosh\phi_1 + \frac{m_2}{m_1})^2 + \sinh^2\phi_1} \,.
\label{v2}
\ee
$\cosh\chi_1$ and $\sinh\chi_1$, and hence $v_1$, can be found using 
eqs.(\ref{rel2momcons}). 

If the initial data are rational, then $\cosh\chi_1$, $\sinh\chi_1$, 
$\cosh\chi_2$ and $\sinh\chi_2$ are all rational. The outgoing velocities 
$v_1$ and $v_2$, and their gamma factors, are therefore rational too.  
The expressions (\ref{chi2}) for $\cosh\chi_2$ and $\sinh\chi_2$ 
have the form (\ref{hypPythpts}) associated with a Pythagorean triple, with 
\be
(M,N) = L \left(\cosh\phi_1 + \frac{m_2}{m_1} \,, \ \sinh\phi_1
\right) \,.
\label{Pythag2out}
\ee
Generally, $\cosh\phi_1 + \frac{m_2}{m_1}$ and $\sinh\phi_1$ are not 
integers, and one must multiply by a common factor $L$ to 
find the integers $(M,N)$ of a primitive triple. 

As an example, suppose $m_1 = 2$ and $m_2 = 1$, and the particles have 
initial velocities $u_1 = \frac{3}{5}$ and $u_2 = 0$. The velocity $u_1$ 
is associated to the Pythagorean triple $(4,3,5)$, with 
$(\cosh\phi_1, \, \sinh\phi_1) = (\frac{5}{4}, \frac{3}{4})$.
From eq.(\ref{Pythag2out}), $(M,N) = (7,3)$. So, from eqs.(\ref{chi2}) and
(\ref{v2}), $(\cosh\chi_2, \, \sinh\chi_2) = (\frac{29}{20}, \frac{21}{20})$,
corresponding to the Pythagorean triple $(20,21,29)$, and $v_2 =
\frac{21}{29}$. Further, from eq.(\ref{rel2momcons}), 
$(\cosh\chi_1, \, \sinh\chi_1) = (\frac{41}{40}, \frac{9}{40})$,
corresponding to the triple $(40,9,41)$, and $v_1 = \frac{9}{41}$.
(Another simple example is with $u_1 = \frac{4}{5}$, $u_2 = 0$ and 
masses $m_1 = 3$ and $m_2 = 1$.)

It is not necessary to assume that particle 2 is initially at rest. 
The outgoing velocities and gamma factors are still rational for any 
rational initial data. This is because if initially $u_1$ and $u_2$ 
are nonzero, we can perform a Lorentz boost to make $u_2$ vanish, 
calculate the result of the collision as above, and then apply the 
inverse boost. The boost, which acts on each 2-momentum in the same 
way, is the rational matrix
\be
\left(\begin{array}{cc}
\cosh\phi_2 &  -\sinh\phi_2 \\
-\sinh\phi_2 & \cosh\phi_2 
\end{array}\right) \,,
\ee
and reversing the sign of $\phi_2$ gives the inverse boost.
So the calculations preserve rationality at each step. More generally, the
notion of rational relativistic collision is invariant under any boost
in the rational subgroup of the Lorentz group SO(1,1).

It is also useful to consider the collision
in the centre of mass frame. From general initial data one finds this 
frame by calculating the total 2-momentum 
\bea
P = (E, p) &=& (m_1\gamma(u_1) + m_2\gamma(u_2), \,
m_1\gamma(u_1)u_1 + m_2\gamma(u_2)u_2) \nonumber \\ 
&=& (m_1\cosh\phi_1 + m_2\cosh\phi_2, \,
m_1\sinh\phi_1 + m_2\sinh\phi_2)
\label{totalEp}
\eea
and then performing the Lorentz boost that makes $p$, the total
spatial momentum, vanish. This boost is
\be
(E^2 - p^2)^{-\half}
\left(\begin{array}{cc}
E &  -p \\
-p & E 
\end{array}\right) \,.
\label{CMboost}
\ee
In the centre of mass frame, the particles just reverse their velocities 
in the collision, so $v_1 = -u_1$ and $v_2 = -u_2$, and $\gamma(v_1) 
= \gamma(u_1)$ and $\gamma(v_2) = \gamma(u_2)$. This solves 
eq.(\ref{relMomcons}). 

Therefore, in general, the outgoing particle 2-velocities are found by boosting 
the incoming 2-velocities to the centre of mass frame, reversing the signs of 
the spatial velocity components, and then inverting the boost. This combined
operation is the improper Lorentz transformation
\bea
&& (E^2 - p^2)^{-1}
\left(\begin{array}{cc}
E & p \\
p & E 
\end{array}\right)
\left(\begin{array}{cc}
1 & 0 \\
0 & -1 
\end{array}\right)
\left(\begin{array}{cc}
E & -p \\
-p & E 
\end{array}\right) \nonumber \\
&& \qquad\qquad\qquad\qquad = (E^2 - p^2)^{-1}
\left(\begin{array}{cc}
E^2 + p^2 & -2Ep \\
2Ep & -(E^2 + p^2) 
\end{array}\right) \,.
\label{boorefl1}
\eea
For general, rational initial data, $E$ and $p$ are rational, and this 
matrix is too, even though the matrix (\ref{CMboost}) is not. The
outgoing 2-velocities are therefore rational. The outgoing 2-velocity
of particle 2 is
\be
\left(\begin{array}{c}
\cosh\chi_2 \\
\sinh\chi_2
\end{array}\right)
= (E^2 - p^2)^{-1}
\left(\begin{array}{cc}
E^2 + p^2 & -2Ep \\
2Ep & -(E^2 + p^2) 
\end{array}\right)
\left(\begin{array}{c}
\cosh\phi_2 \\
\sinh\phi_2
\end{array}\right) \,,
\label{boorefl2}
\ee
and similarly for particle 1. The outgoing velocity of particle 2 is
\be
v_2 = \tanh\chi_2 = \frac{2Ep - (E^2 + p^2)\, u_2}
{(E^2 + p^2) - 2Ep \, u_2} \,,
\label{boorefl3}
\ee
which agrees with eq.(\ref{v2}) when $u_2 = 0$. $v_1$ is given by the same
formula, with $u_2$ replaced by $u_1$. Note that this does not mean
that $v_2$ is independent of $u_1$, nor $v_1$ of $u_2$, because $E$
and $p$ depend on both $u_1$ and $u_2$.

\vspace{4mm} 

\section{Collision as a 2-velocity reflection}
\vspace{3mm}

In the centre of mass frame, the elastic collision of two particles 
reverses the signs of both spatial velocities. This is 
a spatial reflection in 2-velocity space. In a general frame of 
reference, the effect of the collision is the same reflection, 
conjugated by the Lorentz boost taking this frame to the centre of 
mass frame.

The conjugated reflection is the matrix (\ref{boorefl1}), which acts 
linearly on 2-velocities as in eq.(\ref{boorefl2}). Note that if a particle 
2-velocity is $U = (U^0, U^1) = (\gamma(u), \gamma(u)u)$ then $u$ is
the inhomogeneous, or projective, variable $U^1/U^0$. That is why the 
conjugated reflection acts on the particle velocity by the fractional 
linear transformation associated to the matrix (\ref{boorefl1}), as we
see in eq.(\ref{boorefl3}). 

To clarify its geometrical properties, let us suppress the particle
label, and write the fractional linear transformation (\ref{boorefl3}) as
\be
v = \frac{2Ep - (E^2 + p^2)\, u}
{(E^2 + p^2) - 2Ep \, u} \,.
\label{boorefl4}
\ee
This is an involution between $u$ and $v$, a transformation whose
square is the identity. Its orbits are pairs of points that are 
exchanged by the involution, together with two fixed points.

Generally, a fractional linear transformation
\be
v = \frac{a - bu}{c - du}
\label{inv1}
\ee
can be rewritten as $a - bu - cv + duv = 0$.
It is an involution if it is symmetric in $u$ and $v$, in other words,
if $b = c$, because the transformation from $u$ to $v$ is then the
same as the transformation from $v$ to $u$. Suppose the involution, 
which is now $a - b(u+v) + duv = 0$, has the further property
of exchanging $u=1$ and $v=-1$. Then $u+v = 0$, $uv = -1$ is a
solution, so $d = a$. The involution simplifies to
\be
a - b(u+v) + auv = 0 \,.
\label{inv3}
\ee
For the transformation (\ref{boorefl4}), the coefficients are such
that it simplifies in this way, with $a = 2Ep$ and $b = E^2 + p^2$. 
The fixed points are found by setting $v=u$. They are 
$u = \frac{p}{E}$ and $u = \frac{E}{p}$. 

The transformation (\ref{boorefl4}) is in fact completely determined
by the conditions that (i) it is an involution, (ii) it maps $-1$ and
$1$ into each other, and (iii) it has the centre of mass velocity
$\frac{p}{E}$ as a fixed point.

Physically, only velocities between $-1$ and $1$ occur. The
involution transforms any velocity $u$ in this range into a velocity
$v$ in the same range. The physically realisable fixed point is 
$u = \frac{p}{E}$, with $|u| < 1$. It describes the situation where 
one incoming particle is moving at the velocity of the centre of
mass. The other incoming particle then has the same 
velocity, and the collision is at negligible relative speed. The 
outgoing particles have essentially the same velocities as the incoming ones.  

Geometrically, an involution, if non-trivial, is always a type of 
reflection, but here we can be explicit, and show
that the transformation (\ref{boorefl1}) is a genuine reflection 
in 2-velocity space.

In the Euclidean plane, a reflection $R$ is defined by 
\be
R(\x) = \x - 2(\x \cdot \n)\n \,,
\ee
where $\n$ is a unit vector, satisfying $\n \cdot \n = 1$. $R(\x)$ is 
the reflection of $\x$ in the line passing through the origin whose 
unit normal is $\n$. $R(\x)$ has the same length as $\x$, and 
$R(R(\x)) = \x$.   

In the 2-velocity plane with Lorentzian metric, let $n$ be a spatial
unit vector, satisfying $n \cdot n = -1$. The reflection defined by
$n$, acting on a 2-velocity $U$, is
\be
R(U) = U + 2(U \cdot n)n \,.
\ee
It is easy to verify that if $U \cdot U = 1$ then $R(U) \cdot R(U) =
1$. Also $R(R(U)) = U$. In component form, the reflection is
\be
\left(\begin{array}{c}
R(U)^0 \\
R(U)^1
\end{array}\right)
=
\left(\begin{array}{cc}
1 + 2(n^0)^2 & -2n^0n^1 \\
2n^0n^1 & 1 - 2(n^1)^2 
\end{array}\right)
\left(\begin{array}{c}
U^0 \\
U^1
\end{array}\right) \,.
\label{2velrefl}
\ee

We require the vector $n$ to be normal to $P = (E, p)$, the total 
2-momentum, in the Lorentzian sense that 
$n \cdot P = n^0 E - n^1 p = 0$. This fixes $n$ to be
\be
n = (E^2 - p^2)^{-\half} \, (p, E) \,,  
\ee
a unit vector that makes the same angle with the spatial momentum axis
as $P$ makes with the energy axis. For this $n$
the matrix in (\ref{2velrefl}) reproduces the matrix 
(\ref{boorefl1}). It follows that if $U$ is the incoming 2-velocity of 
either of the particles that collide, then its reflection $V = R(U)$ is the
outgoing 2-velocity of the particle. 

A simple way to describe the effect of the collision is therefore to
decompose the incoming 2-momenta as 
\be
m_1 U_1 = \alpha \, P + \beta \, n \,, \quad 
m_2 U_2 = (1 -\alpha) \, P - \beta \, n \,,
\ee
using $P$ and $n$ as basis 2-vectors. The outgoing 2-momenta are then 
\be
m_1 V_1 = \alpha \, P - \beta \, n \,, \quad 
m_2 V_2 = (1 -\alpha) \, P + \beta \, n \,.
\ee
Total 2-momentum is conserved, and each particle's 2-momentum remains
on its mass-shell, because $n \cdot P = 0$. 

This finally leads to a geometrical picture of the effect of the 
collision in the 2-momentum plane. The incoming 2-momenta, $m_1 U_1$ 
and $m_2 U_2$, are on the mass-shell hyperbolae for particles of 
masses $m_1$ and $m_2$. The total 2-momentum 
$P$ is the vector sum of these. The outgoing 2-momenta, $m_1 V_1$ 
and $m_2 V_2$, are on the same mass-shells and are the reflections 
of the incoming 2-momenta in the line through the origin, $O$, and $P$. 
These reflections are in the direction $n$, the Lorentzian normal 
to $OP$. The 2-momenta are shifted to the opposite ends of diameters 
of the hyperbolae parallel to $n$. These shifts are of equal magnitude 
but opposite sense, because $P$ is the total momentum, and 
because the line $OP$ bisects these diameters \cite{Som}. 
A typical collision is illustrated in Figure 1.

\begin{figure}[ht]
\begin{center}
\leavevmode
\vskip -0cm
\includegraphics[width=\textwidth]{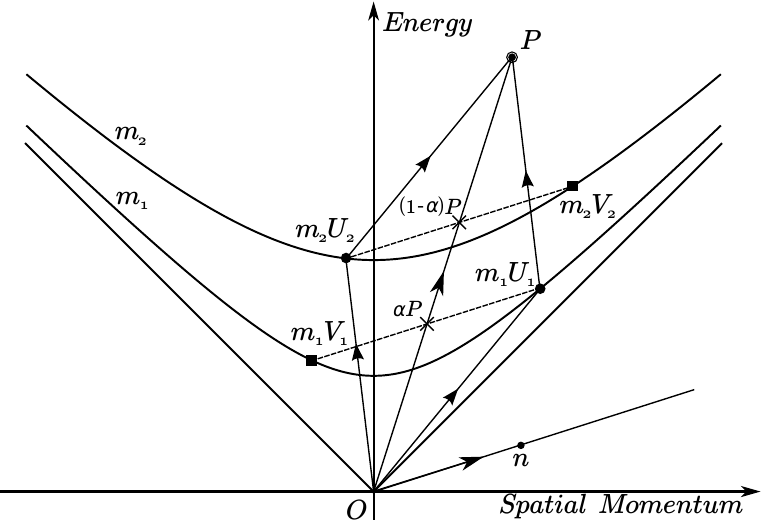}
\caption{2-momentum diagram for a relativistic, two-particle elastic
collision. The outgoing 2-momenta $m_1V_1$ and $m_2V_2$ are obtained by
a Lorentzian reflection of the incoming 2-momenta $m_1U_1$ and $m_2U_2$ 
in the line through $O$ and the total 2-momentum $P$. 
The hyperbolae are the $m_1$ and $m_2$ mass-shells. $n$ is the 
Lorentzian unit normal to $OP$. $\alpha$ is the fraction of $P$ 
carried by the first particle.}
\vskip 0cm
\end{center} 
\end{figure}

\vspace{4mm}

\section*{Acknowledgements}

I am most grateful to Gary Gibbons, Nuno Rom\~ao and David Tong for helpful 
comments, to Jerrold Franklin for correspondence, and to Chris Lau for
producing the figure.

This work was partly carried out during the author's visit to Tours
for the Le Studium conference on Gravitation, Solitons and Symmetries.

\end{document}